\documentclass[a4paper,preprint]{aastex}

\usepackage{natbib}

\shorttitle{HAT-P-7b secondary eclipse measurements}
\shortauthors{Christiansen et al.}

\begin{document}

\title{Studying the atmosphere of the exoplanet HAT-P-7b via secondary eclipse measurements with \textit{\textbf{EPOXI}}, \textit{\textbf{Spitzer}} and \textit{\textbf{Kepler}}}

\author{Jessie L. Christiansen\altaffilmark{1}, Sarah Ballard\altaffilmark{1}, David Charbonneau\altaffilmark{1}, N.~Madhusudhan\altaffilmark{2}, Sara Seager\altaffilmark{2}, Matthew J. Holman\altaffilmark{1}, Dennis D. Wellnitz\altaffilmark{3}, Drake Deming\altaffilmark{3}, Michael F. A'Hearn\altaffilmark{3} and the {\it EPOXI} team}

\altaffiltext{1}{Harvard-Smithsonian Center for Astrophysics, 60 Garden Street, Cambridge, MA 02138; jchristi@cfa.harvard.edu}
\altaffiltext{2}{Massachusetts Institute of Technology, Cambridge, MA 02159}
\altaffiltext{3}{University of Maryland, College Park, MD 20742}

\begin{abstract}
The highly irradiated transiting exoplanet, HAT-P-7b, currently provides one of the best opportunities for studying planetary emission in the optical and infrared wavelengths. We observe six near-consecutive secondary eclipses of HAT-P-7b at optical wavelengths with the {\it EPOXI} spacecraft. We place an upper limit on the relative eclipse depth of 0.055\% (95\% confidence). We also analyze {\it Spitzer} observations of the same target in the infrared, obtaining secondary eclipse depths of $0.098\pm0.017$\%, $0.159\pm0.022$\%, $0.245\pm0.031$\% and $0.225\pm0.052$\% in the 3.6, 4.5, 5.8 and 8.0 micron IRAC bands respectively. We combine these measurements with the recently published {\it Kepler} secondary eclipse measurement, and generate atmospheric models for the dayside of the planet that are consistent with both the optical and infrared measurements. The data are best fit by models with a temperature inversion, as expected from the high incident flux. The models predict a low optical albedo of $\lesssim0.13$, with subsolar abundances of Na, K, TiO and VO. We also find that the best fitting models predict that 10\% of the absorbed stellar flux is redistributed to the night side of the planet, which is qualitatively consistent with the inefficient day-night redistribution apparent in the {\it Kepler} phase curve. Models without thermal inversions fit the data only at the 1.25-$\sigma$ level, and also require an overabundance of methane, which is not expected in the very hot atmosphere of HAT-P-7b. We also analyze the eight transits of HAT-P-7b present in the {\it EPOXI} dataset and improve the constraints on the system parameters, finding a period of $P = 2.2047308\pm0.0000025$~days, a stellar radius of $R_{\star} = 1.824\pm0.089 R_{\odot}$, a planetary radius of $R_p = 1.342\pm0.068 R_{\rm Jup}$ and an inclination of $i = 85.7^{+3.5}_{-2.2}$~deg.

\end{abstract}


\section{Introduction}

The characterization of exoplanetary atmospheres is a quickly maturing field. Thus far, observations have revealed an intriguing range of results that are in turn feeding back into a rapidly evolving theoretical paradigm. The most fruitful method of investigation to date is the detection of secondary eclipses of hot Jupiters---highly irradiated gas giant planets in short period (2--3 day) orbits. By measuring the relative depths of the secondary eclipse of an exoplanet in multiple bandpasses, one can reconstruct a low-resolution emission spectrum from the dayside of the planet.

The planet/star flux ratio is highest at infrared wavelengths, due to the thermal emission from the planet. The {\it Spitzer Space Telescope} has been used to measure the infrared secondary eclipse depths for a growing list of extrasolar planets. These measurements reveal that the atmospheres span two distinct categories: those with temperature inversions in the upper atmospheres (HD~209458b, \citealt{Knutson08a}; XO-1b, \citealt{Machalek08}; TrES-4, \citealt{Knutson09}; and XO-2b, \citealt{Machalek09}), and those without (TrES-1, \citealt{Charbonneau05}; HD 189733b, \citealt{Charbonneau08}). It has been proposed that the distinguishing factor between the two classes is the incident flux on the planet, with higher incident flux resulting in gas-phase TiO/VO \citep{Fortney08, Burrows08} or non-equilibrium products of photochemistry, as seen in solar system bodies \citep{Burrows08,Zahnle09}. The additional absorber in the upper atmosphere generates the temperature inversion. Orbiting an F6 star, with a large stellar radius and high stellar temperature, in a short orbital period, HAT-P-7b is one of the most highly irradiated hot Jupiters, receiving $4.5\times10^9$~erg~s$^{-1}$~cm$^{-2}$. It is therefore predicted to have a temperature inversion.

At optical wavelengths, the planet/star flux ratio is an order of magnitude less favorable than in the infrared, due to the much lower thermal emission from the planet at the shorter wavelengths, and also the low optical reflectance expected from cloud-free atmosphere models. Although observing at optical wavelengths makes the detection of the secondary eclipse more difficult, it can provide very useful information. For instance, in the case of HD 209458b, the observed temperature inversion implies the presence of an unknown absorber high in the atmosphere, as mentioned above \citep{Burrows07}. Subsequently, \cite{Rowe08} placed very stringent upper limits on the optical depth of the secondary eclipse using the {\it Microvariablity and Oscillations of Stars} ({\it MOST}) satellite, and concluded that the absorber must have low reflectance. Optical secondary eclipses have been detected for CoRoT-1b \citep{Snellen09} and CoRoT-2b \citep{Alonso09} with relatively low significance using the {\it CoRoT} spacecraft, and with high significance for HAT-P-7b \citep{Borucki09} using the {\it Kepler} spacecraft. The only detection of an optical secondary eclipse from the ground at the time of submission is for OGLE-TR-56b in the $z^{\prime}$ band \citep{Sing09}.

The next step is to combine the optical and infrared secondary eclipse measurements to extend the planetary emission spectrum and refine statements about the atmospheric properties. Prior to the launch of {\it Kepler}, observations of HAT-P-7b were obtained with the EPOCh (Extrasolar Planet Observation and Characterization) component of the NASA {\it EPOXI} Mission of Opportunity. One of the science goals of this mission was to measure or place meaningful upper limits upon the geometric albedo of a small set of transiting extrasolar planets. The goal of this paper is to combine the HAT-P-7b secondary eclipse measurements made with the {\it EPOXI}, {\it Kepler} and {\it Spitzer} spacecraft to generate a broadband spectrum covering $0.35-8\mu$m, and to identify a set of model planetary atmospheres that reproduce the observations. The paper is organized as follows: In Section \ref{sec:epoxi}, we describe the {\it EPOXI} observations and data analysis; in Section \ref{sec:spitzer}, we describe our analysis of the {\it Spitzer} observations of HAT-P-7; in Section \ref{sec:combined} we combine the results of these two data sets with the published {\it Kepler} secondary eclipse measurement and describe the model fits to the final spectrum; and in Section \ref{sec:future} we discuss the conclusions of the paper and the extension of this analysis to additional {\it EPOXI} targets.

\section{{\it EPOXI} observations and analysis}
\label{sec:epoxi}

HAT-P-7b was the final transiting extrasolar planet target observed with {\it EPOXI}. We observed eight transits: the first in the `pre-look', an initial two day observation window (UT 2008 August 8 and 9) to refine the pointing, the latter seven observed consecutively in a fifteen day observation run lasting from UT 2008 August 16 to August 31. Seven secondary eclipses were observed during the run. We obtained 26,781 observations of HAT-P-7 in total.

We observed HAT-P-7 with the 30-cm diameter High Resolution Instrument (HRI) on-board the {\it EPOXI} spacecraft. We used the 1024$\times$1024 pixel CCD with a clear filter, providing wavelength coverage from 350 to 1000 nm. The instrument is significantly defocused, with the PSF having a FWHM of 9.0 pixels (3.7 arcsec). Due to on-board memory limiting the number of full frame images that could be stored, we observed in two sub-array modes: 128$\times$128 pixels typically, and 256$\times$256 pixels during times of select transits and eclipses to minimize pointing losses. Observations were taken nearly continuously, with gaps of approximately 7 hours every 2--3 days to download the data from the spacecraft. 

The data reduction pipeline is detailed in \citet{Ballard09} and summarized here. We receive calibrated data products from the existing Deep Impact pipeline \citep{Klaasen05}, to which we apply an additional bias correction. We use PSF-fitting to find the position of the target star in each image to within a hundredth of a pixel; we discard 4850 images at this stage due to the star falling out of the field of view or too close to the edge of the CCD for accurate photometry. We discard an additional 40 images due to significant (15-$\sigma$) deviations from the model PSF, caused by readout smear or radiation events. We then apply several corrections based on our knowledge of the CCD architecture, including scaling down the flux values in the two rows and two columns that form the internal boundaries at the center of the CCD, which has readout electronics in each of the four corners. We also scale the counts in images taken in 256$\times$256 mode to those taken in 128$\times$128 mode, to account for a systematic flux offset between the two modes. Finally, we divide the images by a flat field generated from a green LED stimulator lamp mounted beside the CCD, which corrects for changes in the position-dependent sensitivity that have occurred since the detector was characterized prior to launch. We obtained these calibration frames (`stims') in six short blocks, each of approximately 20 minutes duration, spaced every few days during the observations of HAT-P-7. Each block contained 250 images, and in total we obtained 500 stims in 128$\times$128 mode and 1000 in 256$\times$256 mode. We correct each science frame in a given mode with the stims obtained in that mode.

\begin{figure}[h!]
\begin{center}
 \includegraphics[width=6.5in]{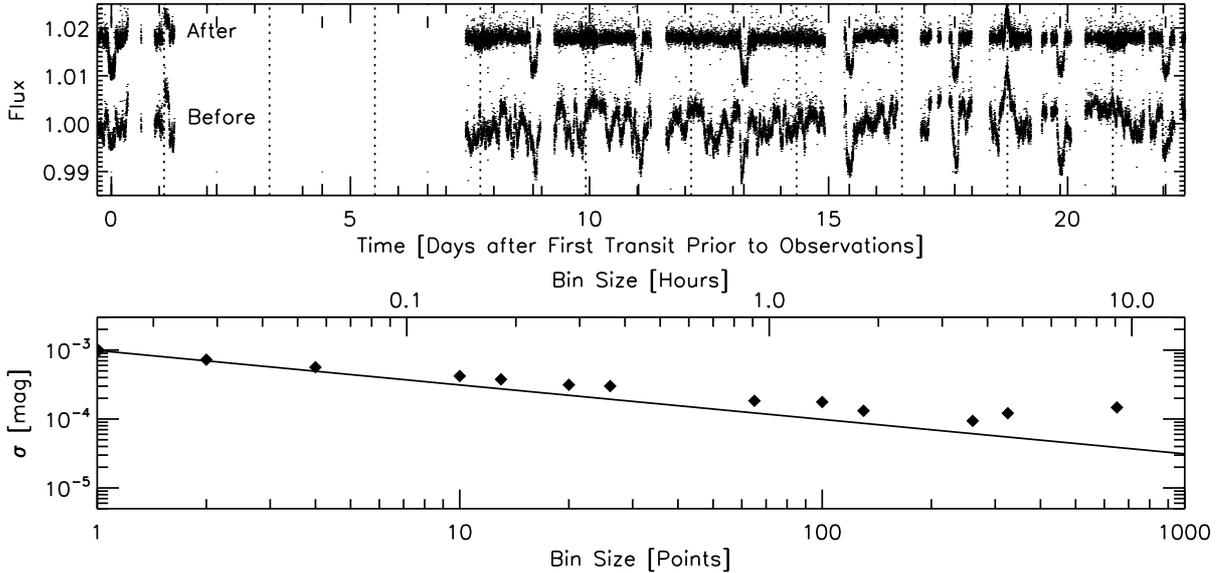}
 \caption{{\it Upper panel}: The full HAT-P-7 {\it EPOXI} light curve. The first transit was observed during the two day `pre-look' that established the pointing; a further seven were observed consecutively over the following 15 day observation run. The lower light curve has not had the surface spline correction applied; the upper light curve includes the correction and is the final light curve used in the analysis. The two brightening events are instrumental in origin (see text). The vertical dashed lines indicate the expected secondary eclipses. {\it Lower panel}: The scatter in the out-of-transit data (shown as diamonds) does not decrease with increasing bin size as expected for Gaussian noise ($1/\sqrt{N}$, where $N$ is the number of points in the bin, shown as the solid line normalized to the unbinned value of the scatter), indicating that there is a component of correlated noise remaining in the light curve.}
   \label{fig:lc}
\end{center}
\end{figure}

We perform aperture photometry on the resulting images, using a 10-pixel radius aperture. At this stage there is still considerable correlated noise in the data, shown as the lower light curve in Figure~\ref{fig:lc}, due to the time-variable jitter in the spacecraft pointing. For the observations of HAT-P-7 this jitter was on the order of 10 arcsec (40 pixels) hr$^{-1}$. We find that the monochromatic green stims are a poor representation of the interpixel sensitivity, which seems to have a color-dependent component, and as a result the pointing jitter introduces correlated noise that we are unable to remove with the calibration frames. We subsequently use the out-of-transit and out-of-eclipse data to map out the brightness variations with CCD position. For a small random subset (6\% of the 21684 points in the light curve) we find an iteratively sigma-clipped mean of the closest spatially neighbouring points, and use these values to generate a surface. We then fit a 2-dimensional spatial spline to the surface, and correct each point in the light curve by interpolating onto this surface. We discard an additional 207 frames affected by readout smear or radiation events at this stage, resulting in a final set of 21684 images. The full corrected light curve is shown as the upper light curve in Figure~\ref{fig:lc}, with an out-of-transit scatter that is 91\% above that expected from Poisson noise. The excess is due to the remaining correlated noise that we have been unable to calibrate, and is higher than the 56\% value we attained for the {\it EPOXI} observations of GJ 436 \citep{Ballard09}. This is due to the fact that the HAT-P-7 light curve has approximately two-thirds the number of observations of GJ 436, and that these observations are distributed over a larger fraction of the CCD. Both of these effects decrease the robustness of the spatial spline and result in higher residual correlated noise.

There are two `flare'-type events visible in the light curve, both of which we believe to be instrumental artifacts. The first occurs during the pre-look, at which time one of the four CCD quadrants was affected by a particular systematic flux correlation that was observed for other {\it EPOXI} targets. For the remainder of the HAT-P-7 observations this systematic effect was not present, and therefore the data obtained during the pre-look in this particular quadrant are not well calibrated by the spline. The second, which can be seen at approximately 18.6 days in Figure~\ref{fig:lc}, occurs at a time when the star was illuminating a part of the CCD to which it does not return in the course of the HAT-P-7 observations. The spline procedure we describe above therefore fails, since it relies on multiple visits to the same parts of the detector.


In order to determine the limb-darkening coefficients across the {\it EPOXI} bandpass, we use the four stellar model atmospheres (Kurucz 1994, 2005) that bracket the HAT-P-7 values of $T_{\rm eff} = 6350$K and log $g = 4.07$, with [M/H] $ = 0.3$ and $v_{\rm turb} = 2.0$ km/s \citep{Pal08}. We found previously that for the wide {\it EPOXI} bandpass the choice of limb-darkening treatment did not systematically affect the results \citep{Ballard09}. We fit the four coefficients of the non-linear limb-darkening law of \citet{Claret00} to 17 positions across the stellar limb. We repeat this fit in 0.2 nm intervals across the 350--1000 nm bandpass, weighted for the total sensitivity (including filter, optics and CCD response) and photon count at each wavelength. We calculate the final set of coefficients as the average of the weighted sum across the bandpass, bi-linearly interpolated between the four models, for which we find $c_1 = 0.43$, $c_2 = 0.14$, $c_3 = 0.46$, and $c_4 = -0.25$.

\subsection{Transits}
\label{sec:transits}

\begin{figure}[h!]
\begin{center}
 \includegraphics[width=5in, angle=90]{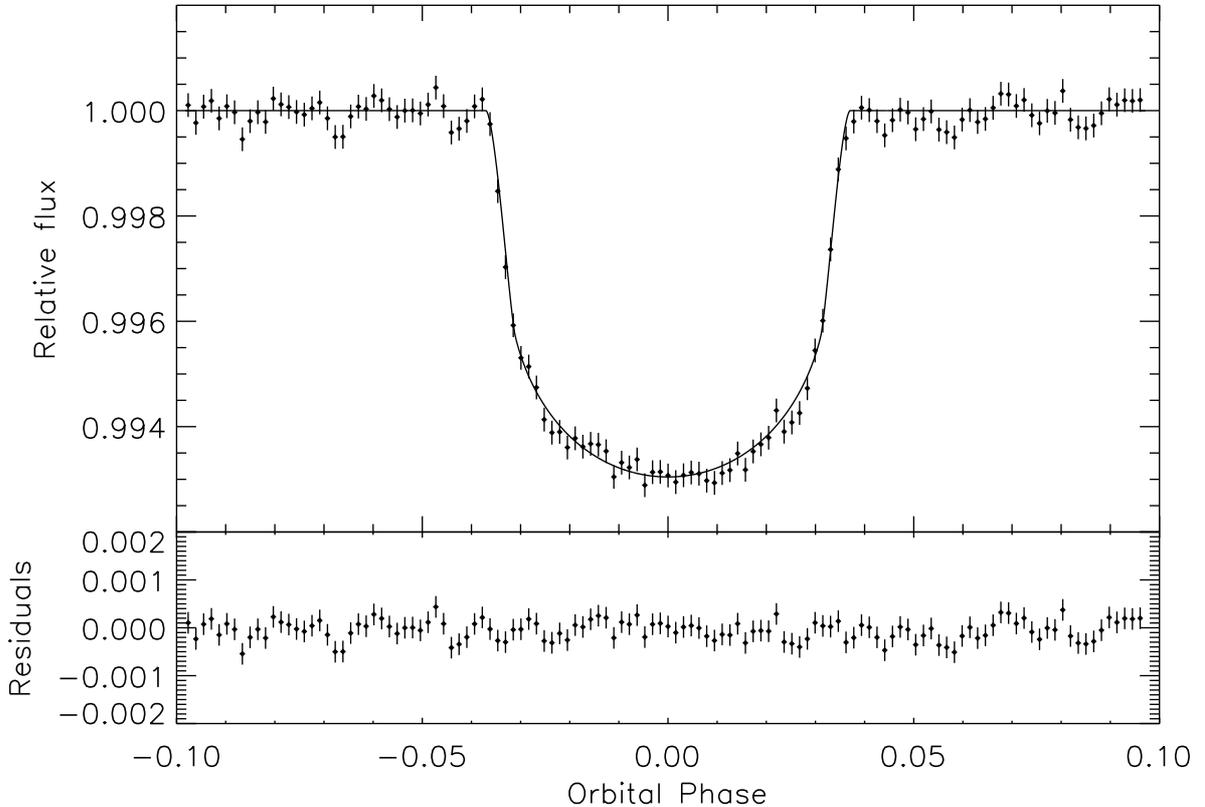}
 \caption{The {\it EPOXI} HAT-P-7 light curve, phase-folded and binned in five minute intervals. The solid line shows the best fit transit model, and the lower panel shows the residuals after this model is subtracted from the data.}
   \label{fig:binnedlc}
\end{center}
\end{figure}

\begin{figure}[h!]
\begin{center}
 \includegraphics[width=3in, angle=90]{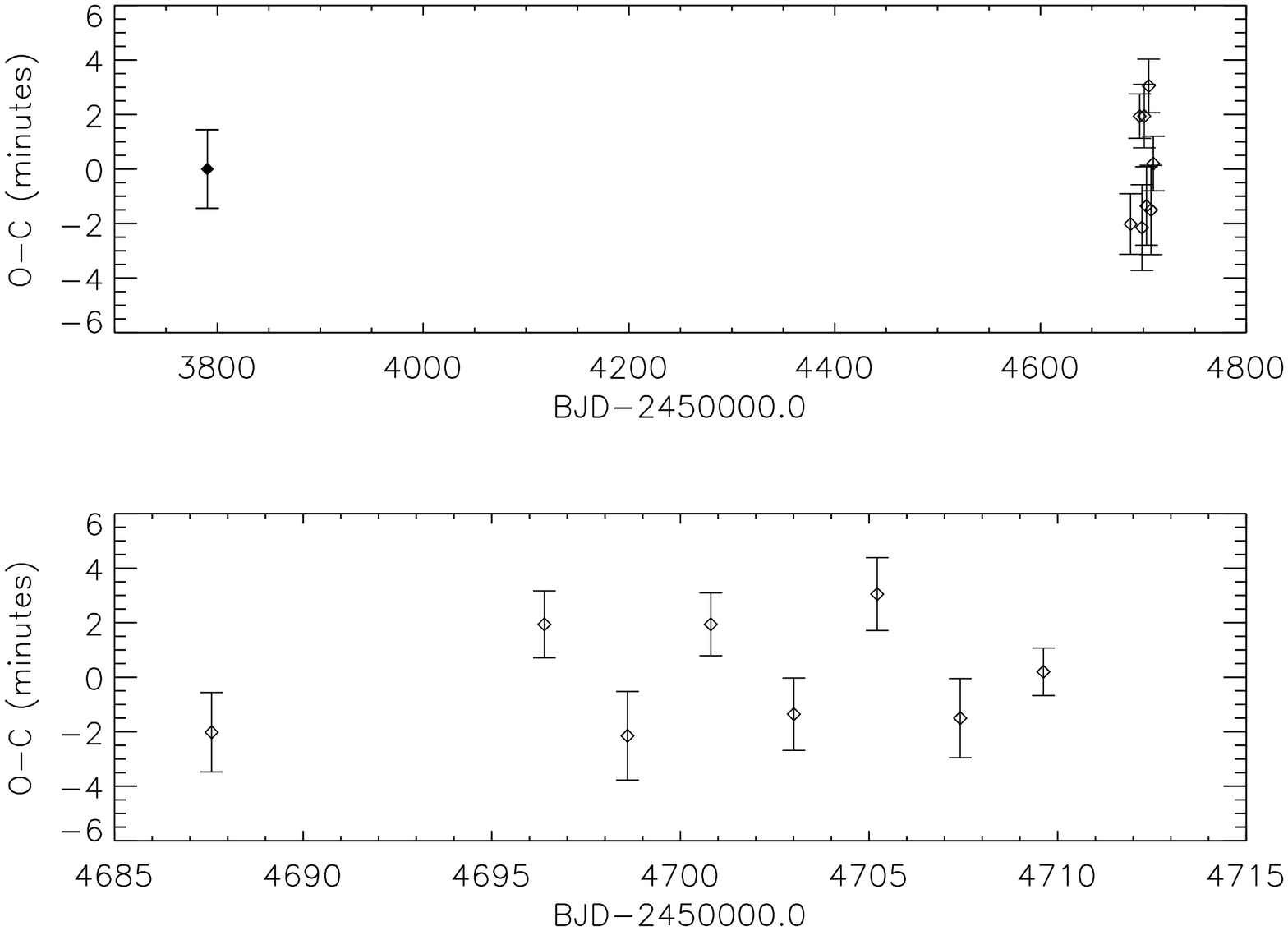}
 \caption{Observed transit times of HAT-P-7b minus the best fit linear ephemeris. {\it Upper panel}: The solid data point is from \cite{Pal08}, the hollow points are the {\it EPOXI} transit times. {\it Lower panel}: An expanded view of the {\it EPOXI} transit times.}
   \label{fig:ttvs}
\end{center}
\end{figure}

We first consider the transits observed by {\it EPOXI}. We fit each transit with a linear, time-dependent slope using four hours of data before and four hours of data after the transit, to remove any residual instrumental linear trend. We use the analytic algorithms of \citet{Mandel02} to fit the transit parameters, using $\chi^2$ as the goodness-of-fit estimator. Using the Levenberg-Marquardt algorithm, we fit for three geometric parameters: $R_p/R_{\star}$, where $R_p$ and $R_{\star}$ are the planetary and stellar radii respectively; $R_{\star}/a$, where $a$ is the semi-major axis; and cos $i$, where $i$ is the inclination of the orbit. We also fit for eight times of transit center for a total of eleven free parameters. We initially fix the period to the value from the discovery paper \citep{Pal08}, 2.2047299~days. We find $R_p/R_{\star}=0.0756\pm0.0013$, $R_{\star}/a = 0.2249\pm0.0116$ and cos $i = 0.0754^{+0.0374}_{-0.0608}$. In order to account for the effects of the remaining systematics in the error budget, we use the ``rosary bead'' residual permutation method described by \citet{Winn08}, with 2000 permutations. The quoted 1-$\sigma$ errors encompass 68\% of the permutations. The light curve is shown in Figure~\ref{fig:binnedlc}, phased and binned in 5 minute intervals, with the best fit model transit overplotted. Using the stellar mass value derived by \citet{Pal08}, $M_* = 1.47\pm0.08 M_\odot$, we find $R_{\star} = 1.824\pm0.089 R_{\odot}$, $R_p = 1.342\pm0.068 R_{\rm Jup}$ and $i = 85.7^{+3.5}_{-2.6}$~deg, which are consistent with and an improvement over the values in the discovery paper. However, they are moderately inconsistent with the parameters presented in the recently posted \citet{Winn09} paper; a rigorous comparison of the results is not undertaken here. The new {\it EPOXI} values and the best fit times of transit center are given in Table \ref{tab:pars}. For the transit times, shown in Figure~\ref{fig:ttvs}, we see no evidence for transit timing variations on the order of 1--2 minutes, and we use a weighted linear fit to derive a new ephemeris combining the discovery ephemeris and the eight times listed here: $T_c \rm{(BJD)} = 2454700.81308\pm0.00100$, $P = 2.2047308\pm0.0000025$~days. 


\begin{deluxetable}{lc}
\tabletypesize{\scriptsize}
\tablecaption{HAT-P-7 system parameters}
\tablewidth{0pt}
\tablehead{\colhead{Parameter} & \colhead{Value}}
\startdata
Adopted values\tablenotemark{a} & \\
$M_*$ $(M_{\odot})$ & $1.47\pm0.08$ \\
$M_p$ $(M_{\rm Jup})$ & $1.776\pm0.077$ \\
\\
Derived values & \\
$P$ (days) & $2.2047308\pm0.0000025$ \\
$T_c$ (BJD) & $2,454,700.81308\pm0.00100$ \\
$R_{\star} (R_{\odot})$ & $1.824\pm0.089$ \\
$R_p (R_{\rm Jup})$ & $1.342\pm0.068$ \\
$i$ (deg) & $85.7^{+3.5}_{-2.2}$ \\
Transit times (BJD) &   $2,454,687.5815\phantom{6}\pm0.0012\phantom{1}$ \\
&  $2,454,696.40413			  \pm 0.00075$\\
&  $2,454,698.6064\phantom{3}			  \pm 0.0012\phantom{3}$\\
&  $2,454,700.8139\phantom{3}			  \pm 0.0011\phantom{6}$\\
&  $2,454,703.01513			  \pm 0.00099$\\
&  $2,454,705.22219			  \pm 0.00086$\\
&  $2,454,707.43077                         \pm 0.00074$\\
&  $2,454,709.63062                         \pm 0.00092$\\
\enddata
\tablenotetext{a}{Masses are from \cite{Pal08}.}
\label{tab:pars}
\end{deluxetable}

\subsection{Secondary eclipses}
\label{sec:eclipses}

\begin{figure}[h!]
\begin{center}
 \includegraphics[angle=90,width=5in]{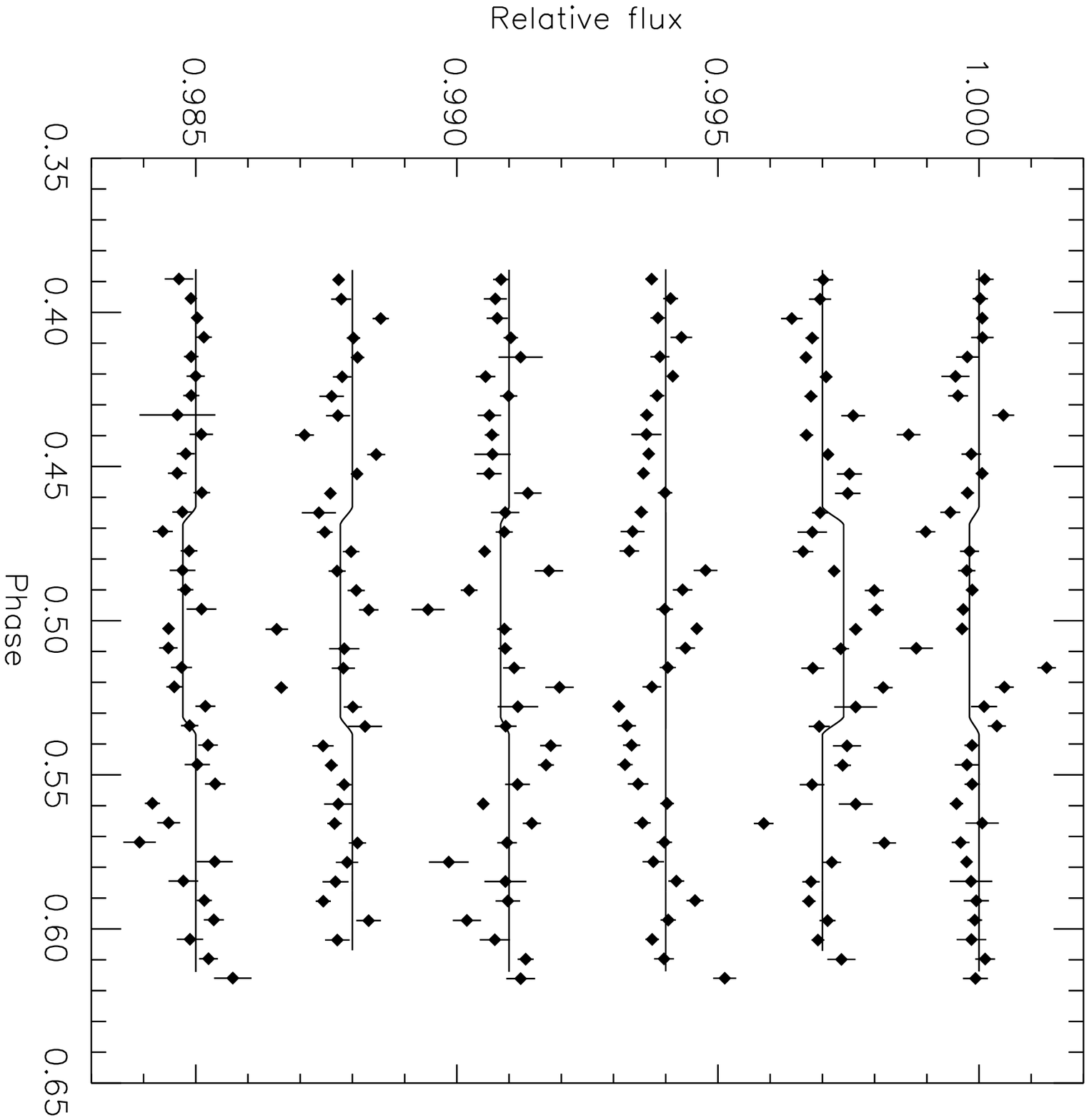}
 \caption{The relative flux versus orbital phase for the six HAT-P-7b eclipses observed with {\it EPOXI}, binned in 20 minute intervals, and offset by a constant for clarity. The solid line shows the best fit eclipse model in each case.}
   \label{fig:eclipses}
\end{center}
\end{figure}

We monitored eight secondary eclipses of HAT-P-7b with {\it EPOXI}. Two of the eclipses are interrupted by the systematic effects described earlier, and for the final analysis, we remove these two eclipses and use the remaining six, shown in Figure~\ref{fig:eclipses}. For this analysis, we perform an additional calibration step. When producing the full HAT-P-7 light curve, we use only 6\% of the data points to create the 2-dimensional spatial spline, in order to minimize suppression of any additional astrophysical signals in the data, such as transits of additional planets in the system. However, when trying to extract the shallow signal of the eclipses of HAT-P-7b, we can make use of all of the data points that are not obtained during those events to increase the robustness of the calibration. For each event, we consider the points that occur within $\pm1.5$ times the transit duration of HAT-P-7b from the predicted center of eclipse, resulting in a window of approximately 12 hours duration. For each point in this window, we find a resistant mean of the fluxes of the nearest 30 data points on the CCD that do not occur during a transit or eclipse of HAT-P-7b, and normalize the flux of the point by this value. This reduced the scatter of the points in each window by 1--2\% on average. Since we find that we are still ultimately limited by systematic errors in the resulting light curves, we use the scatter in the six measured depths as a proxy for our precision. 

For each eclipse, we bin the data in 20 minute intervals, rejecting 3-$\sigma$ outliers within each bin. As with the transit analysis, we fit the data on either side of the eclipse with a linear slope with time to remove longer timescale trends. We generate a model using the geometric parameters derived from the transit fitting and with no limb-darkening, and scale the model depth to that which minimizes the $\chi^2$ of the fit to the binned data. We do not allow the time of secondary eclipse to vary, fixing it at the expected phase value of 0.5, since there is no evidence of eccentricity for HAT-P-7b (see Section \ref{sec:spitzer}). We use the rosary bead method to determine the 1-$\sigma$ error bars, shifting the residuals and re-binning each time, with approximately 820 permutations per eclipse. The six eclipse depths are given in Table \ref{tab:epoxi}, where a negative eclipse depth corresponds to an increase in flux at the predicted time of secondary eclipse. The best fit eclipse models are overplotted in Figure~\ref{fig:eclipses}. The error bars in the figure are calculated from the scatter in each bin normalized by the square-root of the number of points in that bin. For the final value we take the mean of the six measurements; the systematics preclude us from scaling down the individual error bars when combining the measurements, resulting in a best fit depth of $0.0070\pm0.0240$\%, where the error is set such that 2-$\sigma$ (0.048\%) encompasses the six points. 

\begin{deluxetable}{cc}
\tabletypesize{\scriptsize}
\tablecaption{HAT-P-7b {\it EPOXI} secondary eclipse measurements. Negative values imply increases in flux at the predicted time of eclipse.}
\tablewidth{0pt}
\tablehead{\colhead{Predicted time of eclipse (BJD)} & \colhead{Eclipse depth}}
\startdata
2,454,695.30127 & $0.018\pm0.014$\% \\
2,454,697.50586 &$-0.041\pm0.011$\% \\
2,454,699.71045 & $0.000\pm0.015$\% \\
2,454,701.91553 & $0.016\pm0.018$\%\\
2,454,704.12026 & $0.023\pm0.010$\%\\
2,454,708.52979 & $0.025\pm0.006$\%\\
Average depth &   $0.0070\pm0.0240$\%\\
\enddata
\label{tab:epoxi}
\end{deluxetable}

\section{{\it Spitzer} observations and analysis}
\label{sec:spitzer}

Secondary eclipses of HAT-P-7b were observed with the {\it Spitzer Space Telescope} \citep{Werner04} under a Target of Opportunity program (program 40685) led by Joseph Harrington. Observations of two consecutive eclipses were obtained with the InfraRed Array Camera (IRAC; \citealt{Fazio04}): the first was observed on UT 2008 October 28 in the 3.6 and 5.8 micron channels for 9.1 hours, the second on UT 2008 October 30 in the 4.5 and 8.0 micron channels for 8.1 hours. The latter were preceded by a 30 minute preflash set of observations, described further in Section~\ref{sec:long}.  We treat the data in the same fashion as \citet{Charbonneau08, Knutson08a, Knutson09, Machalek08, Machalek09}. We summarize the procedure here, with particular care taken to describe any ways in which our analysis differs from earlier work.

For each channel, we perform aperture photometry on the BCD (Basic Calibrated Data) products, using an aperture radius of 2.0 pixels for the 3.6 and 5.8 micron channels, 2.5 pixels for the 4.5 micron channel, and 3.0 pixels for the 8.0 micron channel. We determine the centers of the apertures by calculating the weighted sum of the flux in a circle of radius 3 pixels positioned at the approximate location of the star. We choose the aperture size to minimize the scatter in the out-of-eclipse data. From this point, we treat the two shorter wavelength channels in a different manner to the two longer wavelengths and the analyses are described separately.

\subsection{3.6 and 4.5 micron channels}

For each image, the position of the aperture was compared to the median of the preceding 10 and following 10 images. Where the position deviated by more than 3 times the rms of these values from the median, the estimate of the centroid was corrupted and the image discarded. For the 3.6 micron channel, 26 (0.53\%) of 4906 images were discarded; for the 4.5 micron channel, 11 (0.25\%) of 4370 images were discarded. For both channels, the sky background was calculated in an annulus centered on the aperture with inner and outer radii of 10 and 15 pixels respectively and subtracted. 

Observations in these shorter wavelength channels are characterized by a position-dependent flux sensitivity \citep{Reach05, Charbonneau05, Knutson08a, Knutson09}; we fit for this sensitivity and the eclipse light curve simultaneously. We generate the light curve with the analytic equations from \citet{Mandel02} with no limb darkening, using the revised HAT-P-7b parameters from the {\it EPOXI} analysis. We model the intra-pixel flux sensitivity for both channels with linear and quadratic coefficients in x and y position, as shown in Equation \ref{eq:intra-pixel}, where the measured flux $f^{\prime}$ is a function of the incident flux $f$ and position; adding higher order position terms or time-dependent terms did not significantly improve the final fit.

\begin{equation}
\label{eq:intra-pixel}
f^{\prime} = f(b_1 + b_2(x - \bar x) + b_3(x - \bar x)^2 + b_4(y - \bar y) + b_5(y - \bar y)^2)
\end{equation}

For each channel, there are seven free parameters in the fit: a constant, a linear and a quadratic coefficient in x position, a linear and quadratic coefficient in y position, an eclipse depth, and a time of center of eclipse. For the fitting, we use the Monte Carlo Markov Chain (MCMC) method \citep{Ford05, Holman06}: in each case we run four chains of 500,000 steps, varying one of the five parameters at each step, and discarding the first 20\% of each chain. The best fit value of each parameter is taken as the median of the distribution, and the error as the range centered on the median that encompasses 68\% of the total points. We find depths of $0.098\pm0.017$\% and $0.159\pm0.022$\% in the 3.6 and 4.5 micron bands respectively. The results are shown in Figures~\ref{fig:uncorr} and \ref{fig:corr}: the former shows the uncorrected data in 6 minute bins, with the model of intra-pixel sensitivity overplotted, the latter the binned data with the sensitivity divided out, and the best fit eclipse model overplotted. In order to assess the effect of any remaining correlated noise in the data we also use the rosary bead method described in Section \ref{sec:transits}. In each case the number of permutations of the residuals is set to the number of points in the light curve (4880 points and 4359 points in the 3.6 and 4.5 micron bands respectively) and the eclipse fit performed with the seven free parameters described above. We measure nearly identical eclipse depths and errors for the rosary bead method as for the MCMC method, finding $0.099\pm0.014$\% in the 3.6 micron band and $0.162\pm0.016$\% in the 4.5 micron band. For the remainder of the paper we adopt the MCMC eclipse depths and errors.

\subsection{5.8 and 8.0 micron channels}
\label{sec:long}

The 5.8 and 8.0 micron channels show a long timescale time-dependent (as compared to position-dependent) flux sensitivity. The correlation is strongest in the 8.0 micron channel, where an increase in the flux during the course of the observations, referred to as the `detector ramp' is evident. One interpretation of this is an increase in the effective gain of the pixels with time in the 8.0 micron detector. One successful method for mitigating this ramp is the preflash technique \citep{Knutson09b}, where a bright diffuse region is observed immediately prior to the start of the target observations in order to pre-emptively increase the effective gain of the pixels. For the 8.0 micron observations of HAT-P-7, a 30 minute set of preflash observations, 3648 images in total, were obtained of the HII region GAL 075.83+00.40 ($\alpha=$ 20 21 39.5, $\delta=$ +37 31 04). The median flux in the pixels in the aperture used for the HAT-P-7 observations ranges from 3500-4500 MJy sr$^-1$. After the pre-flash a time-dependent correlation is evident in the data, albeit significantly attenuated when compared with similar time series analysis performed in this channel (for example, \citet{Deming07}). As noted by \citet{Charbonneau08} and \citet{Knutson08a} we find that removing the most significantly affected data, in this case the first 18 minutes (79 of 2185 images, 3.6\% of the total), removes the need for higher order terms in the fit to the correlation described below. We also perform an iterative 3-$\sigma$ flux cut over the running median to discard outliers, possibly caused by radiation events, which removes 47 (1.9\%) of 2453 images in the 5.8 micron channel, and 51 (2.4\%) of the remaining 2106 images in the 8.0 micron channels.

We chose the time-dependent function that minimized the out-of-eclipse scatter, which was a linear function of ln($dt$) in both channels, where $dt$ is the time elapsed from the center of the first observed transit, shown in Equation \ref{eq:ln_dt}. Previous analyses have used quadratic functions of ln($dt$) \citep{Charbonneau08, Knutson08a, Knutson09, Machalek08, Machalek09}, however by comparison the ramp is attenuated in both channels in these data, and adding higher order functions of time or position-dependent functions did not improve the fit.

\begin{equation}
\label{eq:ln_dt}
f^{\prime} = f (c_1 + c_2({\rm ln}(dt + 0.02)))
\end{equation}

\begin{figure}[h!]
\begin{center}
 \includegraphics[width=5in]{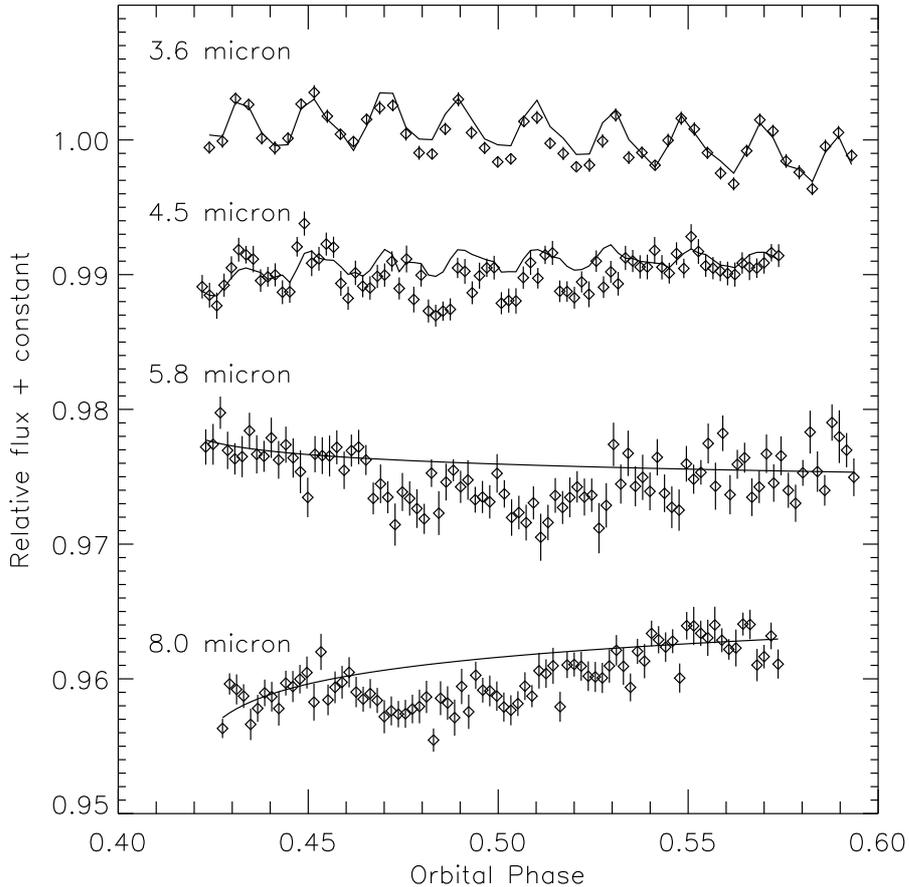}
 \caption{Secondary eclipses of HAT-P-7b observed in the four {\it Spitzer} IRAC bands. The data are binned in 6 minute intervals and offset in flux for clarity. The solid line is the best fit model of the position- and time-dependent instrument effects in each case, described in Section \ref{sec:spitzer}.}
   \label{fig:uncorr}
\end{center}
\end{figure}

\begin{figure}[h!]
\begin{center}
 \includegraphics[width=5in]{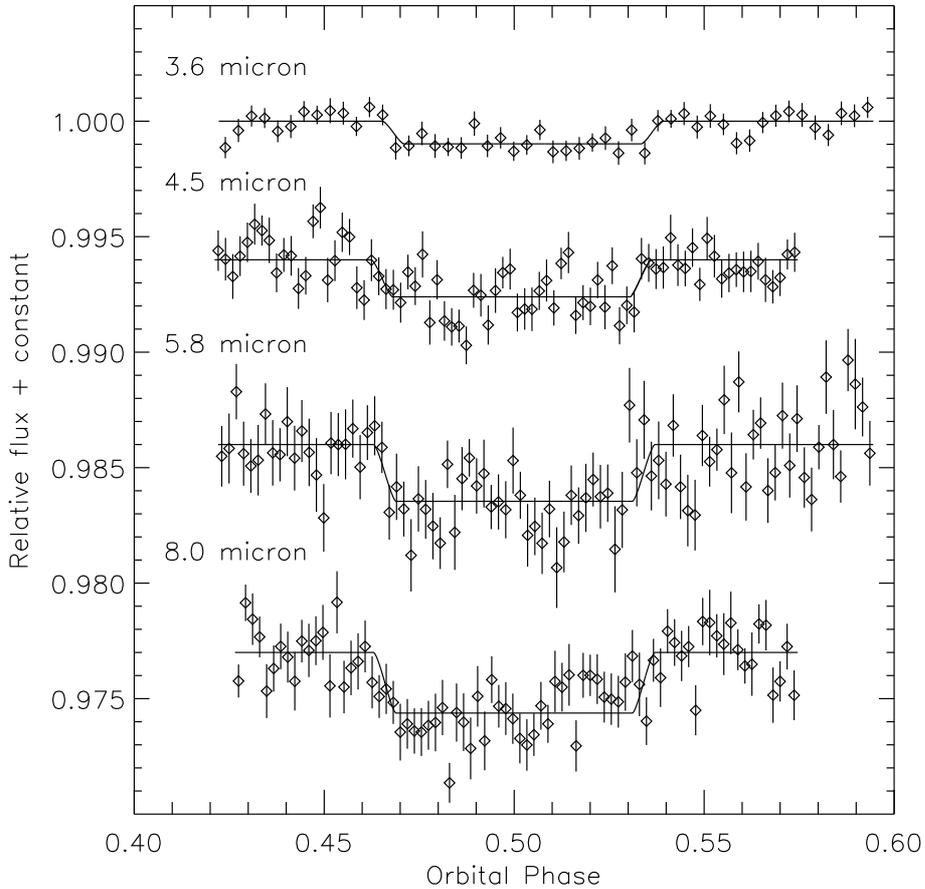}
 \caption{The same data as in Figure~\ref{fig:uncorr}, with the position- and time-dependent correlations removed. Overplotted is the best fit eclipse model from the MCMC analysis in each case. The best fit depths and times of eclipse are given in Table \ref{tab:depths}.}
   \label{fig:corr}
\end{center}
\end{figure}

The additional factor of 0.02 prevents the singularity at $dt=0$, and similarly to \citet{Knutson09} we find no significant change in the measured depth of the eclipse on varying this value. The MCMC fit was performed as described previously, with three free parameters in each case: a constant, a coefficient in time, and an eclipse depth. We found that when the time of eclipse was included as a free parameter there was a correlation between the time of eclipse and the coefficient in time. Since the measured times of eclipses in the shorter wavelengths do not deviate significantly from the predicted times for $e=0$, we do not allow the time of eclipse to vary in the final fit for the 5.8 and 8.0 micron channels. We instead fix them to the values predicted by the {\it EPOXI} ephemeris for zero eccentricity. We calculate the values and errors in the same fashion as for the 3.6 and 4.5 micron channels, finding $0.245\pm0.031$\% and $0.262\pm0.027$\% in the 5.8 and 8.0 micron channels respectively. In order to confirm the validity of fixing the time of eclipse, we repeated the analysis, this time drawing the time of eclipse from a Gaussian distribution defined by the times of eclipses and errors measured in the simultaneously observed channels, using the 3.6 micron values for the 5.8 micron channel and the 4.5 micron values for the 8.0 micron channel. We find near-identical results using this analysis. The results are shown in Figures~\ref{fig:uncorr} and \ref{fig:corr}. Again, we performed a rosary bead analysis to measure the effect of remaining correlated noise on the eclipse parameters. We find that a depth of $0.256\pm0.029$\% in the 5.8 micron band, consistent with the depth measuring using the MCMC method. In the 8.0 micron band the depth is $0.225\pm0.052$\%, approximately 1-$\sigma$ shallower and with a larger error than that produced by the MCMC method, indicating that correlated noise represents the dominant contribution to the error budget for this channel. Therefore we adopt the rosary bead method eclipse depth and error for the 8.0 micron band for the following analysis.

\begin{deluxetable}{lccccc}
\tabletypesize{\scriptsize}
\tablecaption{HAT-P-7b {\it Spitzer} IRAC secondary eclipse measurements}
\tablewidth{0pt}
\tablehead{\colhead{Wavelength ($\mu$m)} & \colhead{Depth (MCMC)} & \colhead{Depth (Rosary)} & \colhead{Time (BJD)} & \colhead{O-C (minutes)} & \colhead{T$_b$ (K)}}
\startdata
3.6 & $0.098\pm0.017$\% & $0.099\pm0.014$\% & $2,454,768.05200\pm0.0035$ & $7.20\pm5.04$ & $2250\pm180$ \\
4.5 & $0.159\pm0.022$\% & $0.162\pm0.016$\% & $2,454,770.26413\pm0.0039$ & $-3.17\pm5.62$ & $2600\pm210$ \\
5.8 & $0.245\pm0.031$\% & $0.256\pm0.029$\% & - & - & $3190\pm280$ \\
8.0 & $0.262\pm0.027$\% & $0.225\pm0.052$\% & - & - & $3100\pm240$ \\
\enddata
\label{tab:depths}
\end{deluxetable}

\section{Discussion}
\label{sec:combined}

\begin{figure}[h!]
\begin{center}
 \includegraphics[width=6in]{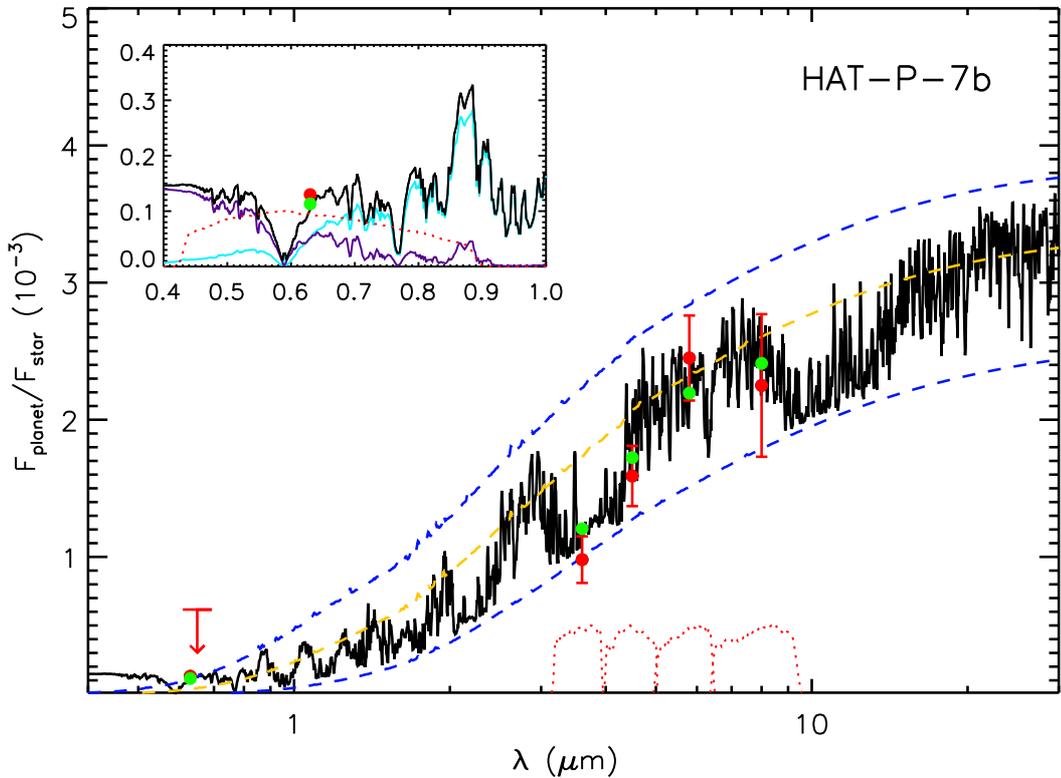}
 \caption{The {\it EPOXI}, {\it Spitzer} and {\it Kepler} secondary eclipse measurements of HAT-P-7b. The solid line is a representative best fit model, with a temperature inversion. The points with error bars are the observed measurements; the points without are the bandpass-integrated model values. The dashed lines show black-body spectra corresponding to 2029~K, 2600~K and 2974~K, for reference. The dotted lines are the instrument response curves. The inset is an expansion of the optical region of the spectrum, showing the {\it Kepler} measurement and response curve. The cyan and purple lines are the thermal and scattered components of the model respectively.}
   \label{fig:spec}
\end{center}
\end{figure}

Figure~\ref{fig:spec} shows the four {\it Spitzer} secondary eclipse depth measurements, the 95\% confidence {\it EPOXI} upper limit at 0.65 micron, and the recently published {\it Kepler} eclipse depth of $0.0130\pm0.0011$\% at 0.63 micron \citep{Borucki09}. This latter measurement was derived from the initial 10 days of {\it Kepler} calibration data and is expected to be refined in the future. We use these eclipse depths to calculate the brightness temperature in each bandpass. Assuming the planet emits as a blackbody, this is a measure of the dayside temperature of the planet required to generate sufficient flux in the bandpass to reproduce the measured planet-to-star flux ratio. In each bandpass, we integrate the ratio of a black body spectrum to a model stellar atmosphere of HAT-P-7, with $T_{\rm eff} = 6500$K, log $g = 4.0$, [M/H] $ = 0.3$ and $v_{\rm turb} = 2.0$ km/s \citep{Kurucz94,Kurucz05}, weighted for the instrument throughput, and solve for the temperature required to match the measured eclipse depth. As expected, the highly irradiated HAT-P-7b has the hottest dayside brightness temperatures measured for any exoplanet, with temperatures in the four bands ranging from 2250K to 3190K. These values are given in Table \ref{tab:depths}. We also calculate this value for the {\it Kepler} secondary eclipse depth measurement \citep{Borucki09} and find $T_{\rm b} = 3175\pm40$K.

We fit the {\it Kepler} and {\it Spitzer} data using the model described in \citet{Madhu09} but extended to visible wavelengths (Madhusudhan \& Seager, in prep.), using a photosphere radius of $1.363R_{\rm Jup}$ \citep{Pal08}. The model assumes global energy balance between the incident and emergent radiation. The HAT-P-7b atmosphere is too hot for condensates; the visible-wavelength opacities include TiO and VO, Na and K, and Rayleigh scattering. We include additional molecular opacities of H$_2$O, CO, CH$_4$, and H$_2$-H$_2$ collision-induced opacities (CO$_2$ and NH$_3$ are included in the model but are not required by the current data). Our H$_2$O, CH$_4$, CO and NH$_3$ molecular line data are from \citep[and references therein]{Freedman08}. Our CO$_2$ data are from R. Freedman (personal communication) and \citet{Rothman05}, and we obtain the H$_2$-H$_2$ collision-induced opacities from \citet{Borysow97}, and \citet{Borysow02}.

\begin{figure}[h!]
\begin{center}
 \includegraphics[width=3in]{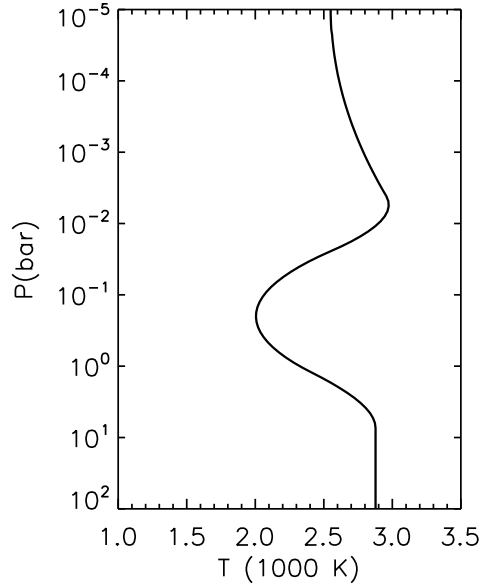}
 \caption{The temperature-pressure profile of HAT-P-7b corresponding to the atmosphere model shown in Figure~\ref{fig:spec}, showing a temperature inversion.}
   \label{fig:pt}
\end{center}
\end{figure}

Our model has fifteen free parameters, and we have at hand only five observational constraints. Consequently, the data allow a large number of solutions at any level of fit. Therefore, we nominally define `best fit' models as those that fit the data to within the 1-$\sigma$ errors shown in Figure~\ref{fig:spec}. An exhaustive exploration of the fifteen dimensional model parameter space is beyond the scope of the present study. Nevertheless, based on strategies outlined in \citet{Madhu09}, we have been able to empirically assess the region in the parameter space that best explains the data. Figure~\ref{fig:spec} shows one representative model that provides a best fit to the observations. The red circles with error bars are the data and the green circles are the bandpass integrated model points. The red dotted lines show the {\it Kepler} and {\it Spitzer} bandpasses. The black curve is the model fit. The blue dashed curves show two black-body spectra corresponding to 2029~K and 2974~K. The orange dashed curve shows a a black-body spectrum at 2600~K, for reference. The inset shows a zoom-in of the {\it Kepler} point and the model in the 0.4-1.0 $\micron$ range. The cyan and purple curves in the inset show the thermal emission and scattered light, respectively. Figure~\ref{fig:pt} shows the pressure-temperature (P-T) profile of the model atmosphere.

We find that all of our best fit models require a thermal inversion to fit the data. A thermal inversion is strongly suggested by the high flux ratio in the 4.5 micron channel of {\it Spitzer} compared to the 3.6 micron channel, in addition to the high flux ratios in the 5.8 and 8 micron channels. The emission features in the 4.5 micron and 5.8 micron channels are due to CO and H$_2$O, respectively. The features in the 3.6 micron and 8 micron channels are predominantly due to CH$_4$. The molecular mixing ratios (with respect to H$_2$) for the model shown in Figure~\ref{fig:spec}, are H$_2$O = 1.0$^{-3}$, CO = 1.0$^{-3}$, CH$_4$ = 1.0$^{-6}$, and no CO$_2$. If we relax the model assumption of thermochemical equilibrium, and allow a `best fit' to the data to be defined by 1.25-$\sigma$ errors, it is possible to generate suitable models that do not have a thermal inversion. However, these models require an overabundance of CH$_4$ in the atmosphere. Given the very hot atmosphere of HAT-P-7b, thermochemical equilibrium favors a high abundance of CO and low CH$_4$. Therefore, a non-inversion model fit to data would imply signs of extreme non-equilibrium chemistry \citep{Madhu09b}.

For a planet as hot as HAT-P-7, both scattered starlight and planet thermal emission could contribute to the planet flux in the {\it Kepler} bandpass \citep{Lopez-Morales07}. The relative amount of scattered starlight and thermal emission depends on the absorbers and scatterers at visible wavelengths: molecular opacities due to TiO and VO, atomic opacities due to Na and K, and Rayleigh scattering mostly from H$_2$. HAT-P-7b is too hot for condensates, and without a reflective condensate layer, visible-wavelength photons will be absorbed before they can be scattered, except at very blue wavelengths. Even without a rigorous quantitative model, therefore, we can infer that the albedo of HAT-P-7b is likely to be low. 

Our best fit models show a low albedo and include the prominent sources of opacity in the optical. The model shown gives a geometric albedo of about 0.13 in the {\it Kepler} bandpass, and has scattered light dominating at wavelengths blueward of approximately 0.6 microns and the planet's thermal emission dominating redward of approximately 0.6 microns. The abundances of Na and K are free parameters in the model and are found to be 0.01 solar; similarly the TiO/VO abundances are found to be $10^{-4}$ of their solar abundances. We note that even though TiO and VO are expected to contribute to the visible and IR opacity, whether they are responsible for the thermal inversion is not known \citep[see][]{Spiegel09}.

We emphasize that the {\it Kepler} data point for HAT-P-7b was instrumental in constraining the P-T profile, beyond the constraints placed by the Spitzer data alone. This is particularly true because of the high planet-star flux ratio of HAT-P-7b; in general, a visible-wavelength observation with a lower contrast may not be as useful for constraining P-T profile.

The values of atomic and molecular abundances present in our models are valid in the framework of a 1D averaged day-side atmosphere. The true values of atomic and molecular abundances, and hence the resulting albedo, in the atmosphere of HAT-P-7b depend on the 3D atmospheric structure, and in 3D are degenerate at the resolution of the {\it Kepler} data point.

The {\it Kepler} phase curve \citep{Borucki09} shows a very inefficient day-night redistribution of absorbed stellar energy. Although energy redistribution is not a focus of this paper, we emphasize that a relatively inefficient day-night redistribution is consistent with, and even required by, our best fit models. We find that approximately 10\% of the incident stellar flux is redistributed to the night side, under the assumption that the visible and infrared thermal phase curves will be similar. 

\section{Conclusions and future work}
\label{sec:future}

We have assembled a broadband emission spectrum of the highly irradiated dayside of HAT-P-7b, extending from 0.35--8.0 micron. We find that this spectrum is best reproduced using an atmospheric model with a temperature inversion, supporting the hypothesis that the presence of an inversion layer is related to the absorbed stellar radiation. A census of exoplanet atmospheres at 3.6 and 4.5 micron that is being undertaken as part of the Warm {\it Spitzer} program will shed more light on the parameter space between where temperature inversions are present in the atmospheres of the most highly irradiated planets and absent in the less irradiated planets. This will include full phase curves of HAT-P-7 at 3.6 and 4.5 microns, which will further constrain the redistribution and energy budget of the planetary atmosphere. In addition, as the {\it Kepler} observations of HAT-P-7 accumulate, the quality of the resulting light curve will be exquisite. This will not only provide higher constraints on the secondary eclipse and phase curve measurements, but will allow investigation into the variability of the planetary emission properties.

At optical wavelengths, we find that the best fit models predict a low geometric albedo of $\lesssim0.13$, which is far less reflective than gas giants in our solar system (e.g. 0.52 for Jupiter). The extremely hot atmosphere precludes the condensation of a reflective layer (i.e. clouds), and our predicted albedo is consistent with those measured for other hot Jupiters \citep{Rowe08}.

The {\it EPOXI} HAT-P-7 secondary eclipse constraint is one of the few opportunities to confirm the {\it Kepler} measurement, given that the relative depth of the signal at visible wavelengths is on the order of $10^{-4}$. We observed a second transiting exoplanet target in the {\it Kepler} field of view with {\it EPOXI}, TrES-2, obtaining similar phase coverage to HAT-P-7, and will present those results in a forthcoming paper. {\it Spitzer} observations of TrES-2 have been obtained in the four IRAC bandpasses, and we envision that an analysis of the extended broadband emission spectrum such as that described in this paper will be similarly fruitful in terms of constraining the properties of the planetary atmosphere. Further {\it EPOXI} targets for which we will provide constraints on the geometric albedo are WASP-3, TrES-3 and HAT-P-4, all of which will have {\it Spitzer} IRAC measurements of the infrared secondary eclipse depths in at least the 3.6 and 4.5 micron channels.

\acknowledgments

We are extremely grateful to the {\it EPOXI}~Flight and Spacecraft Teams that made these difficult observations possible.  At the Jet Propulsion Laboratory, the Flight Team has included M. Abrahamson, B. Abu-Ata, A.-R. Behrozi, S. Bhaskaran, W. Blume, M. Carmichael, S. Collins, J. Diehl, T. Duxbury, K. Ellers, J. Fleener, K. Fong, A. Hewitt, D. Isla, J. Jai, B. Kennedy, K. Klassen, G. LaBorde, T. Larson, Y. Lee, T. Lungu, N. Mainland, E. Martinez, L. Montanez, P. Morgan, R. Mukai, A. Nakata, J. Neelon, W. Owen, J. Pinner, G. Razo Jr., R. Rieber, K. Rockwell, A. Romero, B. Semenov, R. Sharrow, B. Smith, R. Smith, L. Su, P. Tay, J. Taylor, R. Torres, B. Toyoshima, H. Uffelman, G. Vernon, T. Wahl, V. Wang, S. Waydo, R. Wing, S. Wissler, G. Yang, K. Yetter, and S. Zadourian.  At Ball Aerospace, the Spacecraft Systems Team has included L. Andreozzi, T. Bank, T. Golden, H. Hallowell, M. Huisjen, R. Lapthorne, T. Quigley, T. Ryan, C. Schira, E. Sholes, J. Valdez, and A. Walsh. We also thank the remainder of the {\it EPOXI} team, including R.~K. Barry, M.~J. Kuchner, T.~A. Livengood and T. Hewagama at the Goddard Space Flight Center, J.~M. Sunshine at the University of Maryland, D. Hampton at the University of Alaska Fairbanks, C. Lisse at the Johns Hopkins University Applied Physics Laboratory, and J. Veverka at Cornell University. 

Support for this work was provided by the {\it EPOXI} Project of the National Aeronautics and Space Administration's Discovery Program via funding to the Goddard Space Flight Center, and to Harvard University via Co-operative Agreement NNX08AB64A, and to the Smithsonian Astrophysical Observatory via Co-operative Agreement NNX08AD05A. This work is also based on observations made with the {\it Spitzer Space Telescope}, which is operated by the Jet Propulsion Laboratory, California Institute of Technology, under contract to NASA. The authors acknowledge and are grateful for the use of publicly available transit modeling routines by Eric Agol and Kaisey Mandel, and also the Levenberg-Marquardt least-squares minimization routine MPFITFUN by Craig Markwardt. The authors also thank H. Knutson and F. Fressin for several useful discussions about {\it Spitzer} data analysis.

\newpage

\clearpage

\end{document}